\documentclass[twocolumn,showpacs,preprintnumbers,amsmath,amssymb]{revtex4}
\usepackage{tabularx,graphicx}

\usepackage{color}
\usepackage{hyperref}
\hypersetup{
    colorlinks=true,
    linkcolor=blue,
    filecolor=blue,      
    urlcolor=blue,
}

\begin{document}

\newcommand{\beq}{\begin{equation}}
\newcommand{\eeq}{\end{equation}}
\newcommand{\beqn}{\begin{eqnarray}}
\newcommand{\eeqn}{\end{eqnarray}}
\newcommand{\bmath}{\begin{subequations}}
\newcommand{\emath}{\end{subequations}}
\newcommand{\bra}[1]{\langle #1|}
\newcommand{\ket}[1]{|#1\rangle}

\title{The Bohr superconductor}
\author{J. E. Hirsch }
\address{Department of Physics, University of California, San Diego,
La Jolla, CA 92093-0319}

\begin{abstract} 
Superconductors have often been described as `giant atoms'. The simplest description of  atoms that heralded their quantum understanding 
was proposed by Bohr in 1913. The Bohr atom starts from some simple assumptions  and deduces  that the angular momentum
of the electron in Bohr orbits is quantized in integer units of $\hbar$.  This remarkable result, which does not appear to be implicit in the assumptions
of the model, can be interpreted as  a `theoretical proof' of the model's 
validity to describe physical reality at some level. Similarly we point out here that from some simple assumptions  it can be deduced that
electrons in superconductors reside in mesoscopic orbits with orbital angular momentum $\hbar/2$. This implies that both in superconductors and in ferromagnets
 the long-range order results from elementary units of identical angular momentum. Similarly to the case of the
Bohr atom we propose that this remarkable result is compelling evidence that this physics, which is not part of  conventional BCS theory, describes physical reality at some level and  heralds a qualitatively new understanding of superconductors.
 \end{abstract}
\maketitle 

\section{introduction}
The Bohr model of the hydrogen atom \cite{bohr1} is a remarkable bridge between classical and quantum descriptions of the physical world. It is deeply rooted
in classical physics and makes one key  `quantum' assumption motivated by blackbody radiation physics, that in itself does not  appear to
contain the 
assumption of orbital angular momentum quantization. Yet orbital angular momentum quantization is deduced mathematically. We argue that this 
remarkable result can be
taken to be `theoretical proof' that the model describes physical reality at some level, and that as a consequence, the fact that the numerical values
of the Bohr radius, Rydberg constant, etc obtained from the model coincide with experimentally measured values may be interpreted as theoretical confirmation that the experiment was correct rather than as experimental confirmation that the theory is correct.

The Bohr atom description of the hydrogen atom, while being neither  complete nor entirely correct, was absolutely essential in the
developments that led to   the current understanding of atoms based on Schr\"odinger's and Dirac's equations. It seems impossible to imagine 
how one would have reached the Schr\"odinger description of   atoms without having the Bohr atom as a stepping stone.
Similarly, we propose  here that to understand superconductors  it is essential to first understand them at the level of the Bohr atom. It should
be noted that superconductors
have often been characterized as ``giant atoms'' in the past \cite{ga1,ga2,ga3}.  

We will show in this paper that within a simple physical description of superconductors based on electrons residing
in `orbits',  it follows that electrons in these orbits necessarily possess orbital angular momentum $\hbar/2$. This result
does not appear to be implicit in the simple assumptions defining the model, just like in the case of the Bohr atom. We argue similarly
that this result should be interpreted as `theoretical proof' that the model describes the physical reality of superconductors  at some level.

Let us start by reviewing  the Bohr atom. The assumptions made are:

(1) Electrons move in stable circular orbits labeled by integers $n=1, 2, 3, ...$. The n-th orbit has radius $r_n$ and the electron in that orbit
has speed $v_n$ and   energy $E_n$, with $E_n>E_m$ for $n>m$.

(2) The standard relations from classical mechanics and electrostatics relate kinetic and potential energy, velocity and radius of the orbit for an electron
of mass $m_e$, charge $e$ moving in the field of an infinitely massive point charge $-Ze$. Contrary to the prediction from classical electromagnetism it is
assumed that the electron does not emit electromagnetic radiation when it moves in a given orbit.

(3) When the electron makes a transition from orbit $n$ to orbit $m$ ($n>m$), electromagnetic radiation of frequency $(E_n-E_m)/h$ is emitted, where
$h$ is  Planck's constant.

(4) The frequency given in (3) for large $n$ and $m=n-1$ is the frequency of rotation of the electron in the n-th orbit. This is the
`correspondence principle'.

From these assumptions it follows that the angular momentum of the electron in the n-th orbit is given by
\beq
\ell_n\equiv m_ev_nr_n=n \hbar
\eeq 
and all the other consequences of the Bohr theory. Note that the right-hand side of Eq. (1) is independent of the values of
$m_e$, $e$ and $Z$. The derivation of this remarkable result from the assumptions (1)-(4) is well known and straightforward and will not be repeated here.

Similarly, for the superconductor we assume that:

(1) Electrons in the ground state of superconductors move in stable circular orbits. 
The radius of these orbits is determined by the fact that superconductors exhibit the Meissner effect.

(2) The superconducting electrons move in a background of positive charge that neutralizes the superfluid negative charge.

(3) The superconductor has cylindrical geometry and the plane of the orbits is  perpendicular to the cylinder axis.

(4) The spin-orbit interaction derived from Dirac's equation describes the interaction of the electron spin with the background positive charge.

We point out that the fact that the superfluid has negative charge is shown experimentally by the London moment \cite{lm} and the
gyromagnetic effect \cite{gyrosc1,gyrosc2}. The reason to restrict ourselves to cylindrical geometry is that the mathematics is simplest. 

We will show that from these assumptions it follows that the angular momentum of the electrons in these orbits is
\beq
\ell=\hbar/2
\eeq
and argue that the fact that this simple relation follows from the basic assumptions is a strong indication that the model reflects real
physics of superconductors.

\section{$2\lambda_L$ orbits}

We will work throughout in a cylindrical geometry as required  by assumption (3).
 To prove Eq. (2), we first show that the radius of the electron orbits must be $2\lambda_L$, with $\lambda_L$ the London penetration depth,
in order for the superconductor to exhibit a Meissner effect. 

The magnetization of a superconducting  cylinder under applied magnetic field $\vec{H}$ is
\beq
\vec{M}=-\frac{\vec{H}}{4\pi}
\eeq
so that the magnetic field in the interior $\vec{B}=\vec{H}+4\pi\vec{M}=0$.  Assuming that $\vec{M}$ results from orbital motion of $n_s$ carriers
per unit volume, each contributing orbital magnetic moment $\vec{\mu}$
\beq
\vec{M}=n_s\vec{\mu}
\eeq
and the relation between orbital angular momentum and magnetic moment for electrons with mass $m_e$ and charge $e$
\beq
\vec{\mu}=\frac{e}{2m_ec}\vec{\ell}
\eeq
it follows that the relation between orbital angular momentum and applied magnetic field is
\beq
\ell=\frac{m_e c}{2\pi n_s e}H .
\eeq
The relation Eq. (5) for superconductors has been verified experimentally \cite{gyrosc1,gyrosc2}.

The speed of electrons in the Meissner current is obtained from London's equation which follows simply from requiring that, in the relation between
velocity $\vec{v}$ and canonical momentum $\vec{p}$  for the superfluid electron in the presence of magnetic vector potential $\vec{A}$,
\beq
m_e\vec{v}=\vec{p}-\frac{e}{c}\vec{A}  ,
\eeq
$\vec{p}=0$ because of `rigidity' of the superfluid wavefunction \cite{tinkham}, from which it follows that
\beq
v=-\frac{e}{m_ec}A=-\frac{e \lambda_L}{m_ec}H .
\eeq
The relation $A=\lambda_L H$ used in the second equality
follows from Stokes' theorem for a cylindrical geometry of radius $R>>\lambda_L$, where $A$ denotes the magnetic vector potential
at the surface. Replacing $H$ in terms of $v$ in Eq. (6) yields
\beq
\ell=\frac{m_e^2c^2}{2\pi n_s e^2}\frac{v}{\lambda_L}  .
\eeq
Finally, using Ampere's law
\beq
\vec{\nabla}\times\vec{H}=\frac{4\pi}{c}n_s e \vec{v}
\eeq
and the relation derived from Eq. (7)
\beq
\vec{\nabla}\times\vec{v}=-\frac{e}{m_ec}\vec{H}\eeq
it follows that \cite{tinkham}
\beq
\frac{1}{\lambda_L^2}=\frac{4\pi n_s e^2}{m_e c^2}
\eeq
and  Eq. (9) yields
\beq
\ell=m_e v (2\lambda_L)
\eeq
for the angular momentum of each electron in its circular orbit, to yield total magnetization $M$ that will cancel the applied magnetic field.
Electrons move in these  orbits with speed $v$, the interior motions cancel out and a surface current flowing within $\lambda_L$ of
the surface results, the Meissner current. From the definition of angular momentum in circular orbits Eq. (1) and Eq. (13) it follows that the radius of
these orbits is $2\lambda_L$. Since $2\lambda_L$ is much larger than typical interelectronic distances it follows that these orbits
are strongly overlapping.

It is possible to derive this result in several other ways, as follows:

(i) The total angular momentum of the Meissner current in a long cylinder of radius $R$ and height $d$ with applied magnetic field parallel to the
cylinder axis can be written in the two equivalent forms
\beq
L=(2\pi R d\lambda_Ln_s)(m_e v R)=(\pi R^2 d n_s) (m_e v (2\lambda_L))
\eeq
where the first form describes the angular momentum of the supercurrent flowing within $\lambda_L$ of the surface, and the second
form describes the angular momentum of all the charge carriers in the bulk in their orbits of radius $2\lambda_L$ \cite{missing}.

(ii) Faraday induction upon application of a magnetic field $\vec{H}$ changes the speed of an electron in an orbit of radius $2\lambda_L$ perpendicular
to $\vec{H}$ by precisely the amount Eq. (8) giving the
speed of the electrons in the Meissner current \cite{copses}.

(iii) The Lorentz force acting on an electron that expands its orbit from a microscopic radius to radius $2\lambda_L$ in the presence of
magnetic field $\vec{H}$  perpendicular to the plane of the orbit  imparts azimuthal
velocity to the electron of precisely Eq. (8) \cite{copses}.

(iv) The Larmor diamagnetic susceptibility for charge carriers of density $n_s$ in orbits of radius $2\lambda_L$ perpendicular to the applied
magnetic field is precisely $-1/4\pi$, describing perfect diamagnetism \cite{missing}.

We have argued elsewhere that (iii) offers a dynamical explanation of the Meissner effect \cite{meissner1,meissner2} within the 
theory of hole superconductivity \cite{holesc}. The orbit expansion is driven by lowering of kinetic energy \cite{kinetic} and is also associated
with negative charge expulsion from the interior to the surface of the superconductor \cite{chargeexp}.

\section{angular momentum quantization}

We have shown in the foregoing that the magnetization of a cylindrical superconducting body that cancels an applied magnetic field $H$ results from 
superconducting electrons in orbits of radius $2\lambda_L$, with azimuthal speed given by Eq. (8). It is natural to conclude that
electrons reside in such orbits even in the absence of an applied magnetic field, as opposed to assuming that the $2\lambda_L$ orbits
are somehow  `created' by the applied field. If so,   the speed Eq. (8) should be interpreted as the
$difference$ in speed for the electron in the  orbit in the presence and absence of the magnetic field, just like for  a diamagnetic atom. 
It should be noted that the hypothesis that superconducting electrons reside in large orbits was made by
several researchers in the pre-BCS era \cite{orbits1,orbits2,orbits3}.

In a diamagnetic atom
angular momentum is quantized, and in particular the component of the
electron angular momentum parallel to the applied magnetic field  is an integer multiple of $\hbar$. 
Hence it is natural to ask whether the angular momentum of electrons in the $2\lambda_L$ orbits in the superconductor  is also quantized.
Indeed we will show in what follows  that the orbital angular momentum for electrons in these orbits is $\hbar/2$. 

The Hamiltonian for an electron including the
spin-orbit interaction derived from the Dirac equation is \cite{bjorken}
\beq
H=\frac{p^2}{2m_e}-\frac{e\hbar}{4m_e^2 c^2} \vec{\sigma}\cdot(\vec{E}\times\vec{p})
\eeq
where $\vec{E}$
 is an electric field. We can write it in terms of a vector potential $\vec{A}_\sigma$ \cite{ac}
 \bmath
 \beq
 H=\frac{1}{2m_e}(\vec{p}-\frac{e}{c} \vec{A}_\sigma)^2
 \eeq
 \beq
 \vec{A}_\sigma=\frac{\hbar}{4 m_e c}\vec{\sigma}\times \vec{E}   .
 \eeq 
 \emath
 Eq. (15) results from Eq. (16) to lowest order in $\vec{A}_\sigma$ if $\vec{\nabla}\times\vec{E}=0$, which is the case in a time-independent situation.
  Just as in the derivation of the Meissner  velocity Eqs. (7), (8), we assume $\vec{p}=0$ because of `rigidity'. and choose the axis of the cylinder as the
  spin quantization axis. The term in brackets in Eq. (16a) is 
 $m_e \vec{v}_\sigma$, hence
 \beq
 \vec{v}_\sigma=-\frac{e\hbar}{4m_e^2c^2}\vec{\sigma}\times\vec{E}
 \eeq
 The background positive charge in which the electron moves has charge density $|e|n_s$ according to assumption (2), which in a cylindrical geometry gives rise
to an electric field
\beq
\vec{E}=-2\pi en_s\vec{r}
\eeq
in the radial direction, and Eq. (17) yields for $r=2\lambda_L$
\beq
\vec{v}_\sigma=\frac{\pi n_s e^2 \hbar}{m_e^2 c^2}\lambda_L\vec{\sigma}\times \hat{n}  .
 \eeq
 Here, $\hat{n}$ is a unit vector in the radial direction, i.e. pointing towards the lateral surface of the cylinder. Using Eq. (12)
 \beq
 \vec{v}_\sigma=\frac{\hbar}{4m_e \lambda_L}\vec{\sigma}\times \hat{n}
 \eeq
 and $\vec{\sigma}$ is perpendicular to $\hat{n}$, 
 so that the magnitude of the angular momentum (which points along the cylinder axis) in the orbit of radius $2\lambda_L$ is
 \beq
 \ell=m_ev_\sigma (2\lambda_L)=\frac{\hbar}{2}
 \eeq
 in agreement with Eq. (2).
 
 An alternative argument is the following \cite{slafes}. The vector potential Eq. (16b) with the electric field given by Eq. (18) can be written as
 \beq
 \vec{A}_\sigma=\frac{\vec{B}_\sigma\times\vec{r}}{2}
 \eeq
 with the `effective magnetic field' $\vec{B}_\sigma$ given by
 \beq
\vec{B}_\sigma=-\frac{\pi n_s e  \hbar}{m_ec} \vec{\sigma}=-2\pi n_s \vec{\mu}
 \eeq
  with $\vec{\mu}=(e\hbar)/(2m_e c)\vec{\sigma}$ the electron magnetic moment, pointing along the cylinder axis, so that $\vec{\nabla}\times \vec{A}_\sigma=\vec{B}_\sigma$ for a spatially uniform $\vec{B}_\sigma$. 
 Similarly, for the real vector potential $\vec{A}$ we have the relation 
 \beq
 \vec{A}=\frac{\vec{B}\times\vec{r}}{2}
 \eeq
 so that $\vec{\nabla}\times \vec{A}=\vec{B}$ for a spatially uniform $\vec{B}$. The Meissner effect can be understood dynamically from the assumption
 that the electron expands its orbit from microscopic radius to radius $2\lambda_L$ in the presence of a uniform external magnetic field $\vec{H}$ 
  that is not affected by the magnetic field generated by other electrons expanding their orbits \cite{sm,meissner2}, and
 exactly  the same is the case for the
 velocity acquired through  the effective magnetic field $\vec{B}_\sigma$ originating in the spin-orbit interaction:
 replacing $H$ by $B_\sigma$  in Eq. (8), the speed that results is given by the magnitude of the velocity Eq. (20). The assumption that the magnetic field affecting one electron is not modified by the magnetic field
 generated by other electrons, and similarly that the spin-orbit field $B_\sigma$ originates in the full electric field Eq. (18) unscreened by
 the superfluid electronic charge, is consistent with the fundamental
 quantum mechanical notion that the Schr\"odinger wavefunction $\Psi(\vec{r})$ does not screen itself \cite{schr,meissner2}.

 The two alternative ways of looking at the Meissner current embodied in Eq. (14)  are entirely analogous to the situation in a ferromagnetic 
 material \cite{purcell}, depicted in Fig. 1. In that case one can equivalently understand the magnetization as arising from a superposition of microscopic current loops throughout
 the material or as a macroscopic (``bound'') surface current, $ \mathcal{J}=cM$ being the current per unit length, that produces the same magnetic field as a real current. The angular momentum associated
 with the magnetization in the ferromagnet originates in electron spin, and hence the gyromagnetic ratio is $e\hbar /m_e c$, a factor of $2$ larger than for the
 superconductor \cite{barnett,gyrosc1,gyrosc2}.    Another  difference
 is that the current loops for the superconductor are ``mesoscopic'' rather than microscopic, and as a consequence the current flows within $\lambda_L$ of
 the surface rather than right at the surface, and the effective surface current is $ \mathcal{J}=J\lambda_L$, with $J$ the current density in the
 surface layer of thickness $\lambda_L$. Despite these differences this point of view highlights a close relationship  between ferromagnetism and superconductivity,
 a concept that was prevalent in the early days of superconductivity \cite{ferro}  but was lost after  conventional BCS  theory was developed.
 It is very remarkable that within this point of view  the angular momentum of the individual carriers contributing their magnetic
 moment to the overall magnetization is $\hbar/2$ {\it for both superconductors and ferromagnets}.

    \begin{figure}
 \resizebox{8.5cm}{!}{\includegraphics[width=6cm]{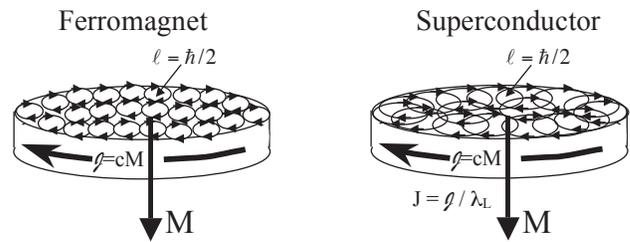}}
 \caption { For a ferromagnet (left panel) the magnetization $M$  can be equivalently understood as arising from a ``bound'' surface current per unit length 
 $ \mathcal{J}=cM$, or from the aggregate of the microscopic orbits depicting the electron spin  in the bulk.
 Similarly for a superconductor (right panel) the magnetization $M$ resulting from an applied magnetic field $H=4\pi M$ in the upward direction can be understood as arising from
 a  current density $J=cM/\lambda_L$ flowing within $\lambda_L$ of the surface or from the aggregate of overlapping mesoscopic orbits of radius
 $2\lambda_L$. For both cases, the angular momentum of each element contributing to the total magnetization is $\hbar/2$.
   }
 \label{figure1}
 \end{figure} 
 
 There are  other arguments that support the conclusion that the angular momentum of electrons in the $2\lambda_L$ orbits is $\hbar/2$, namely:

 (i) We can think of a `Cooper pair' in the absence of applied magnetic field as composed of electrons of opposite spin orbiting
 in opposite direction with speed $v_\sigma$ given by Eq. (20).  When a magnetic field
 perpendicular to the orbits is applied, the Lorentz force on electrons of opposite spin is in opposite directions, $and$ the gradient force on the electron magnetic moment
 due to the varying magnetic field near the surface is in direction opposite to the Lorentz force for both spin orientations. As a consequence,
 the net force is identical for both spin orientations $provided$ the velocity  is given by exactly Eq. (20) (with opposite sign, see \cite{sm} and
 next section), hence the magnetic field will not
 break Cooper pairs apart.
 
 (ii) The magnetic field necessary to impart the electrons  the speed Eq. (20) is, from Eq. (8)
 \beq
 H=-\frac{\hbar c}{4e\lambda_L^2}
 \eeq
 which is precisely the lower critical field $H_{c1}$ of type II superconductors (except for a numerical factor of order 1) \cite{tinkham}. This implies
 that the magnetic field can penetrate the superconductor just at the point when the zero-point motion of the electrons being slowed down by the applied magnetic
 field is stopped \cite{sm}. The magnetization of the system when the magnetic field is $H_{c1}$ is precisely the same that would 
 result from the allignment of all the intrinsic magnetic moments of the superfluid electrons \cite{100,condon}.
 
 \section{the Bohr radius}
 The Bohr radius $a_0=\hbar^2/m_e e^2$ can be understood as resulting from minimizing the total energy
 \beq
 E=\frac{p^2}{2m_e} -\frac{e^2}{r}
 \eeq
 under the constraint that the angular momentum is $\hbar$, hence that the momentum in Eq. (26) is $p=\hbar/r$. Similarly,  consider the Hamiltonian Eq. (16) with
 the electric field given by Eq. (18). It can be written as
 \beq
H=\frac{p^2}{2m_e}+\frac{\hbar}{2m_e}\frac{r}{(2\lambda_L)^2}(\vec{\sigma} \times\hat{n})\cdot\vec{p}
+\frac{\hbar^2}{8m_e}\frac{r^2}{(2\lambda_L)^4} |\vec{\sigma}\times\hat{n}|^4 .
\eeq
using Eq. (12). Under the constraint that the angular momentum is $\hbar/2$, hence $p=\hbar/2r$, the second term in Eq. (27) is a constant independent of $r$ ($\equiv C$), and the energy
is
\beq
E=\frac{\hbar^2}{8m_er^2}
+\frac{\hbar^2}{8m_e}\frac{r^2}{(2\lambda_L)^4} |\vec{\sigma}\times\hat{n}|^4  +C
\eeq
and minimization with respect to $r$ yields   $r=2\lambda_L$, the radius of the orbits in the `Bohr superconductor'.

Note that this constitutes an independent confirmation of the remarkable  consistency of our formalism. For any other value of the radius of the orbit, say $r_0$, the speed Eq. (20) would be
$\hbar /(4m_e\lambda_L)(r_0/(2\lambda_L)$,  the     angular momentum Eq. 
(21) would be $\ell=(\hbar/2)(r_0/2\lambda_L)^2$, (for example $\ell=\hbar$ for $r_0=2\sqrt{2}\lambda_L$), and substituting the momentum $p=\hbar /2r (r_0/2\lambda_L)^2$ in Eq. (27) and minimizing would yield
$r=\sqrt{2\lambda_L r_0}$, different from $r_0$ unless $r_0=2\lambda_L$.

The quadratic term in $A_\sigma$   in the Hamiltonian Eq. (16a), which we derived from  Eq. (15)
by `completing the square', can be physically understood as describing the electrostatic energy cost  that results from the charge expulsion
associated with expansion of the orbits to radius $r$, as discussed in  Ref. \cite{slafes} (Eqs. (19)-(21)). The optimal radius   
$r=2\lambda_L$ results
from the balance between kinetic energy decrease and potential (electrostatic) energy increase as the radius increases, just as in the case
of the Bohr atom.

 \section{the sign problem}

There is a subtle question of $sign$ that needs to be discussed. The effective magnetic field $\vec{B}_\sigma$ 
Eq. (23) is parallel to the spin and will impart a radially  outgoing electron with spin pointing down a clockwise azimuthal velocity, hence an orbital angular momentum
parallel to its spin, as given by Eq. (20). This is incorrect \cite{sm}. In the microscopic atom the spin-orbit interaction gives lowest energy for antiparallel spin and orbital angular 
momentum, and the same should be the case for our macroscopic atom.

In fact, the sign on the right-hand side of Eq. (23) should be $opposite$ if $\vec{\sigma}$ denotes the $electron$ spin, and the same applies to
Eq. (20). The sign is correct provided
$\vec{\sigma}$ is the direction of a $hole$ spin. Recall that the physics we are discussing applies to electronic energy bands that are almost full \cite{holesc},
and $n_s$ in our equations denotes the density of $hole$ $carriers$, not electron carriers \cite{joseph}. For $\vec{\sigma}$ denoting the spin of an electron
and $\vec{v}_\sigma$ the velocity of the electron the correct
form of Eq. (20) is 
 \beq
 \vec{v}_\sigma= - \frac{\hbar}{4m_e \lambda_L}\vec{\sigma}\times \hat{n}
 \eeq
 and the correct form for the effective magnetic field exerting a Lorentz force on the outgoing electron is instead of Eq. (23) 
  \beq
\vec{B}_\sigma=\frac{\pi n_s e  \hbar}{m_ec} \vec{\sigma}=2\pi n_s \vec{\mu}
 \eeq
with $\vec{\mu}$ denoting the electron magnetic moment.

The reader may object that our treatment assumed that carriers had charge $e$, i.e. were electrons ($e<0$ in our notation), hence $\vec{\sigma}$ in
Eqs. (20) and (23) should correspond to the electron rather that the hole spin. In fact however the correct procedure is  to assume that carriers move under the influence of fields as if they were $electrons$ as far as their charge is concerned,
but as $holes$ as far as their spin is concerned. How is that possible?

To understand this rather subtle question (whose answer had eluded us for several years) it is helpful to
recall the  crafty   discussion in Ashcroft and Mermin (AM) \cite{am} on the concept of holes. AM point out that under applied fields both the full and the empty states
in the band evolve as if they were occupied by electrons with negative charge. Because states at the top of the band have negative effective mass, the 
evolution of those states can be equivalently understood as resulting from positively charged electrons with positive mass, i.e. holes. In addition, the spin
associated with the $absence$ of an electron is in opposite direction to that of the electron, so the hole has opposite charge and spin direction
as the electron. In appearance this would not solve our sign problem because Eq. (23) is unchanged if we reverse the sign of both charge and spin.

The remarkable answer is: the sign problem is resolved because  the holes in the superconducting state
are very peculiar objects. Unlike holes in the normal state,
{\it they reside at the bottom rather than the top of the band}  \cite{reorg,reorg2}, hence have $positive$ effective mass. As a consequence they
behave as electrons as far as their charge is concerned and the direction of the Lorentz force due to a magnetic field is concerned,
but as holes as far as their spin is concerned and the direction of the spin-orbit force due to the electric field Eq. (18) is concerned.  Eqs. (26) and (27) are correct
with $e$ denoting $electron$ charge, $\vec{\sigma}$ denoting $electron$ spin, and $n_s$ denoting $hole$ carrier  density,
and the orbital angular momentum is opposite to the spin angular momentum, just as in the microscopic atom.

\section{discussion}
 
We have shown that under very simple and general assumptions it follows that the ground state of a superconductor can be
understood as composed of electrons in mesoscopic orbits of radius $2\lambda_L$, with orbital angular momentum $\hbar/2$. 
In a cylindrical geometry the   orbits are in planes perpendicular to the
cylinder axis, with spin (down) up electrons orbiting (counter)clockwise  in the absence of applied magnetic field,
as   determined by the spin-orbit interaction with the compensating positive charge.  This gives rise to a macroscopic spin
current near the surface \cite{electrospin}, a kind of macroscopic zero-point motion of the superfluid. We propose that this description of superconductors is analogous to the Bohr description of    hydrogen atoms with the electrons in  circular
orbits of angular momentum $n\hbar$, and that the facts that both in our case and in the Bohr atom case angular momentum quantization is derived 
 from very simple assumptions is compelling evidence that these descriptions reflect reality.  
 
 The fact that the orbital angular momentum in these orbits in the superconductor  is  found to be a {\it half integer} rather than an integer multiple of $\hbar$ is very remarkable.
The correct interpretation of this finding could have profound implications. We have suggested that it indicates an intrinsic
double-valuedness of the electron wavefunction \cite{double,double2}, in contradiction with conventional quantum mechanics. Other less radical
interpretations may be possible.

Furthermore we have shown in other work that the expansion of the orbits to radius $2\lambda_L$  is also associated with expulsion of negative charge from the
interior to the surface \cite{chargeexp}, so that the ground  state of the `Bohr superconductor' has excess positive charge near the center and excess negative
charge near the surface, just like the Bohr atom.

Bohr's  correspondence principle played a key role in the formulation of the Bohr atom \cite{bohr1}. It requires that microscopic laws of physics, however
unfamiliar, morph smoothly into the macroscopic laws of physics familiar in our everyday world. It also plays a preeminent role in our
conception of superconductivity. Within BCS theory the explanation of the expulsion of magnetic fields from the interior of superconductors lies
entirely beyond classical physics and hence classical intuition.  Instead, our explanation of the Meissner effect \cite{meissner1,meissner2} is completely consistent with the familiar fact that magnetic field lines
resist motion across a good conductor, so that to expel magnetic field lines it is helpful to also expel  charge.

The Bohr atom description of the hydrogen atom is of course not correct. The correct description (at the non-relativistic level)
is provided by the wavefunctions derived from the Schr\"odinger equation. Nevertheless it is remarkable that the Bohr atom reflects a large part of
the true physics of the hydrogen atom, in particular  the energy of the n-th level is correctly given,  the most probable radius for the electron in the orbit of
maximal angular momentum for quantum number $n$ is correctly given, the z-component of angular momentum is $n\hbar$,   etc. Similarly we are  
not proposing that electrons in $2\lambda_L$ orbits with angular momentum $\hbar/2$ give a complete description of the
superconducting ground state. What we are proposing  is that the correct wavefunction of the superconductor, when it is found, will 
necessarily show
physical properties consistent with the picture provided by quantized $2\lambda_L$ orbits with angular momentum $\hbar/2$ and the
associated macroscopically inhomogeneous charge distribution \cite{electrospin}. The 
BCS wavefunction \cite{bcs} does not.

\acknowledgements
The author is grateful to Z. Fisk for a stimulating discussion.

    \end{document}